\documentclass[useAMS,usenatbib]{mn2e}
\usepackage{subfigure}
\usepackage{setspace}
\usepackage{graphicx}

\def\spose#1{\hbox to 0pt{#1\hss}}
\def\simlt{\mathrel{\spose{\lower 3pt\hbox{$\mathchar"218$}}
     \raise 2.0pt\hbox{$\mathchar"13C$}}}
\def\simgt{\mathrel{\spose{\lower 3pt\hbox{$\mathchar"218$}}
     \raise 2.0pt\hbox{$\mathchar"13E$}}}

\title[Luminosity Functions of Galaxy Groups from Semi-Analytic Models]{A 
Comparison of Galaxy Group Luminosity Functions from Semi-Analytic Models}

\author[O.~N. Snaith et~al.]{Owain N. Snaith$^{1}$, Brad K. Gibson$^{1}$, 
	Chris B. Brook$^{1}$, St\'ephanie Courty$^{1,2}$,\newauthor Patricia 
	S\'anchez-Bl\'azquez$^{1,3,5}$, 
	Daisuke Kawata$^4$, Alexander Knebe$^5$ and Laura V. Sales$^6$ \\
        $^1$University of Central Lancashire, Jeremiah Horrocks
            Institute, 
	    Preston, PR1~2HE, United Kingdom\\
        $^2$Centre de Recherche Astrophysique de Lyon, Ecole Normale
            Superieure de Lyon, Lyon, F-69007, France \\
	$^3$Instituto de Astrofisica de Canarias, Via Lactea S/N. E-38205, 
	    La Laguna, Tenerife, Spain\\
	$^4$Mullard Space Science Laboratory, University College London,
	    Holmbury St. Mary, RH5~6NT, United Kingdom\\
	$^5$Universidad Aut\'onoma de Madrid, Deptartamento de F\'isica 
	    Te\'orica, E-28049 Madrid, Spain\\
	$^6$Kapteyn Astronomical Institute, P.O. Box 800, Groningen, The
	    Netherlands}

\begin{document}

\date{\today}
\pagerange{\pageref{firstpage}--\pageref{lastpage}} \pubyear{2008}

\maketitle

\begin{abstract}
Semi-analytic models (SAMs) are currently one of the primary tools with which to model statistically significant ensembles of galaxies. The underlying physical prescriptions inherent to each SAM are, in many cases, different from one another. Several SAMs have been applied to the dark matter merger trees extracted from the Millennium Run, including those associated with the well-known Munich and Durham lineages.  We compare the predicted luminosity distributions of galaxy groups using four publicly available SAMs \citep{DeLucia2006,Bower2006,Bertone2007,Font2008}, in order to explore a galactic environment in which the models have not been explored to the same degree as they have in the field or in rich clusters. We identify a characteristic ``wiggle'' in the group galaxy luminosity function generated using the De~Lucia et~al. (2006) SAM, that is not present in the Durham-based models, consistent to some degree with observations. However, a comparison between conditional luminosity functions of groups between the models and observations of Yang et al. (2007) suggest that neither model is a particularly good match. The luminosity function wiggle is interpreted as the result of the two-mode AGN feedback implementation used in the Munich models, which itself results in flattened magnitude gap distribution. An associated analysis of the magnitude gap distribution between first- and second-ranked group galaxies shows that while the Durham models yield distributions with approximately equal luminosity first- and second-ranked galaxies, in agreement with observations, the De~Lucia et~al. models favours the scenario in which the second-ranked galaxy is approximately one magnitude fainter than the primary,especially when the dynamic range of the mock data is limited to 3 magnitudes. 
\end{abstract}


\section{Introduction}

In an ideal world, simulating and analysing the formation and evolution of galaxies within Gpc-scale cosmological volumes would be accomplished via the use of self-consistent gravitational N-body and hydrodynamical models. While such an approach is feasible in certain restricted situations, it remains impractical for most applications.  Instead, the compromise most widely adopted is based upon the use of Semi-Analytic Models (SAMs).  The merger and assembly histories of galaxies within SAMs are underpinned by high-resolution cosmological N-body simulations, at the ``cost'' of employing \it a posteriori \rm ``semi-analytical'' treatments of the associated baryonic physics. 

In all the SAM models explored in this paper galaxy properties are derived using a range of gas infall, radiative cooling, re-ionisation, AGN and supernovae feedback, morphological transformation, dust and spectrophotometry prescriptions. In general, the inclusion of AGN feedback within the SAM reduces the luminosity and stellar mass of the brightest galaxies. Supernovae feedback is effective in low mass galaxies, where it becomes an important mechanism by which galactic winds are driven and star formation is quenched. Thus, supernova feedback leads to a reduction in the number of low luminosity galaxies. Upon merging, the stars and cold gas of a satellite galaxy (an accreting galaxy) are added to the reservoir of the central galaxy (called `Centrals', henceforth) of the parent halo. SAMs  have reproduced a range of galactic observables, including colours, luminosities, and mass functions.

SAMs come in several flavours, and, although the codes share many similar features as outlined above, they also differ in the way in which certain processes, relating to  baryonic physics, are implemented (e.g., treatment of supernovae and  AGN feedback). These lead different SAMs to produce different solutions to the problem of galaxy formation. Some of these differences have been explored at length in the literature, via direct comparison with both empirical \it field \rm and \it cluster \rm galaxy luminosity functions, which are the {\it extrema} of galaxy environments \citep{Hatton2003, Mo2004, Gonzalez2005, Bower2006, Croton2006}. For example, \citet{Mateus2008} found that the  \citet{Bower2006} and  \citet{DeLucia2006} models give different trends for the temporal evolution of galaxy merger rates based on close pair counting.  \citet{DiazGimenez2008} suggest that the \citet{Bower2006} and  \citet{DeLucia2006}  models reach different conclusions regarding the rate of chance alignment in low velocity dispersion compact groups.  Recent examples of problems encountered by the SAM approach include the excess of low-mass red galaxies, as identified by \citet{Weinmann2006}, and \citet{Baldry2006}. A comprehensive review of the SAM approach can be found in \citet{Baugh2006}.

What has {\it not} been explored, thus far, at least in any formal sense, is the impact of these baryonic physics prescriptions upon the resulting luminosity and stellar mass functions for the most {\it common} of environments, that of galaxy groups. It is to this aim that our current study is focused.  Galaxy groups are environments where galactic evolution is happening at a high rate due to the low velocity dispersion of groups. This means means that galaxy-galaxy interactions are more likely than in clusters. In this paper, we examine the outputs of  four widely-used SAMs applied to the Millennium Run,\footnote{The simulation was carried out by the Virgo Supercomputing Consortium at the Computing Centre of the Max Planck Society in Garching and recovered from \tt http://galaxy-catalogue.dur.ac.uk:8080/Millennium/.\rm} \citet{Springel2005}, in order to quantify the impact of baryonic physics prescriptions upon the resulting compact and loose group luminosity functions. Two of the   models which we examine will be collectively referred to as the  ``Durham models", being those of  \citet{Bower2006}, (D\_B06), and \citet{Font2008}  (D\_F08), which is an updated version of D\_B06, with a more sophisticated treatment of ram pressure stripping. We also analyse two ``Munich models", being those of   \citet{DeLucia2006} (M\_D06 hereafter), and \citet{Bertone2007} (M\_B07), which differs from M\_D06 mainly in the  supernova feedback recipes. A related model by \citet{Croton2006}, of which M\_D06 is a direct descendant, is also referred to in our study.

After outlining our galaxy group cataloguing procedure, constructed using a classical friends-of-friends approach (\S~2), we examine systematically the predicted distributions of luminosity, and first-to-second rank magnitude gap for both compact and loose groups of galaxies, for each of the SAMs under consideration (\S~3). We analyse the luminosity distribution of galaxy groups in the different models so that the next generation of SAMs can improve the implementation of galaxy formation physics.

\section{Models}

The SAMs used in our analysis employ the merger trees associated with the Millennium Simulation \citep{Springel2005} a large N-body simulation corresponding to a significant volume of the visible Universe, and generated using the WMAP Year~1 cosmology \citep{Spergel2003}.\footnote{($\Omega_{m}$,$\Omega_{b}$,$\Omega_{\Lambda}$,h,n,$\sigma_8$)= (0.25,0.045,0.75,0.73,1,0.9)} The simulation used $2160^3$ particles in a periodic box of side length 500$h^{-1}$~Mpc, gravitational softening of 5h$^{-1}$kpc, and individual particle masses of 8.6$\times$10$^8$~M$_\odot$; 64 outputs exist within the Millennium database, ranging from redshift $z$=127 to $z$=0. The simulation was post-processed using a Friends-of-Friends (FoF) algorithm \citep{Geller1983}, in order to identify density peaks, and then SUBFIND \citep{Springel2001} was employed to identify substructure and split spuriously joined haloes. This information was then used to build merger trees for the dark matter haloes, onto which the SAMs are ``mapped''.

Before embarking on a discussion of the predicted luminosity functions resulting from the use of the aforementioned SAMs applied to the Millennium merger trees, it is important to summarise briefly the defining characteristics associated with each of the primary SAMs employed here. We will highlight the different ways in which the codes create merger trees, the way in which galaxy positions are defined, the implementation of satellite disruption and accretion, and the way in which supernova and AGN feedback are implemented.   

\subsection{Durham models \citep{Bower2006, Font2008}}

In the Durham models, merger trees are produced in a manner which follows that of \citet{Helly2003}, and the properties of these trees are described in \citet{Harker2006}.  These models account for ostensibly separate haloes which are joined by a bridge of dark matter and hence can be erroneously put in a single halo by FoF algorithms, and also account for haloes which are only temporarily joined. Accounting for these effects  results in a halo catalogue containing more haloes than in the original FoF catalogue. The merger trees are then constructed from these catalogues by following subhaloes from early times to late times. We note that the merger trees were constructed independently of those in \citet{Springel2005}.

The merging of galaxies, and lifetime of satellite galaxies, are derived using the method presented in \citet{Benson2002}, which is considerably more sophisticated than the method used in \citet{Cole2000}. When dark haloes merge, a new combined dark halo is formed, and the largest of the galaxies contained within is assumed to be the central galaxy, whilst all other galaxies within the halo are satellites. Satellites are then evolved  under the combined effects of dynamical friction, and tidal stripping. These effects are modelled analytically. The initial orbital energy, and angular momentum, of the satellite upon merging are specified. The orbital energy is set using a constant value of $r_c(E)/R_{vir}$ = 0.5, representative of the median binding energy of satellites, while the orbital ellipticity, is chosen to be between 0.1 and 1.0 at random. Given these parameters, the apocentric distance is found, and the orbit equations are integrated at that point. The host and satellite haloes are all assumed to have NFW profiles, while galaxies are modelled as a disc plus spheroid. The satellite galaxies plus halo are then  advanced by calculating the combined  gravitation forces of the host and satellite haloes, as well as the effects of dynamical friction, calculated using Chandrasekhar's formula. The code keeps track of tidal stripping, to remove mass from the satellites. The new halo mass is then used for the next iteration of the orbit equations. 

This integration of the orbital equations continues until one of three conditions are met: (i) the final redshift, (ii) the host merges with a new larger halo, in which case the satellite becomes a part of the new halo and has new orbital parameters assigned, or (iii) the satellite merges with the central galaxy. In this last case, merging takes place when the orbital radius falls below $R_{merge}$ which is the sum of the half mass radius of the central and satellite galaxies. The mass of the merged satellite is then added to the central galaxy. 
The merging times match largely match those of \citet{Cole2000}, however  some satellites have very long merging times, as they loose a great deal of mass through tidal stripping, meaning that dynamical friction forces become very weak.

The Durham models relate supernova reheating directly to the circular velocity of the galaxy disk according to \citet{Cole2000}: 

\begin{equation}
 \dot{M}_{reheat} \propto V_{disk}^{-2}\dot{M}_* ,
\end{equation}

\noindent where $\dot{M}_{reheat}$ is the rate of change of mass of reheated gas, $V_{disk}$ is the disk circular velocity,  $\dot{M}_*$ is the time derivative of stellar mass and $\dot{M}_{eject}$ is the change in the mass of ejected gas. In haloes with a shallow potential, this has the effect of reducing the amount of cold gas available to form stars, by heating the gas back into the hot gas reservoir. The hot gas is dominated by ejection for low mass haloes, and by reheating without ejection for large haloes. For low mass galaxies, supernova feedback is an important mechanism by which galactic winds are driven and star formation is quenched. 

The Durham models implement AGN feedback is such a way as to regulate the cooling of hot gas.  In large haloes with large Eddington Luminosity, the AGN feedback is assumed to balance heating and cooling, thus truncating star formation. This prevents the formation of over-luminous galaxies. While feedback is active, the black hole is assumed to grow proportionally to the cooling luminosity, and gas is accreted due to disk instabilities. The model assumes quasi-hydrostatic cooling for AGN active galaxies, and has a strict transition between AGN ``active'' and ``inactive'' phases. AGN feedback becomes efficient in galaxies of mass greater than $\sim$2$\times 10 ^{11}h^{-1}$~M$_{\odot}$.  

The essential difference in the D\_B06 and D\_F08 models is the implementation of ram-pressure stripping of the hot gas. In the D\_B06 model, along with both Munich models, the hot gas is instantaneously stripped when it enters a halo already containing a central galaxy. In the D\_F08 model this process happens gradually and depends on the orbit of the galaxy. This has the effect of reducing the population of faint red galaxies.

\subsection{Munich models \citep{DeLucia2006, Bertone2007}}
The Munich merger trees \citep{Springel2005} used in  M\_D06 and M\_B07 follow the positions of subhaloes for as long as they can be identified. Identification is possible only when the number of particles bound to a subhalo exceeds the minimum number of particles set by SUBFIND. The trees are constructed by following the most bound halo particles and searching for the descendant halo in the next output.

One of the key differences between the Munich models and those of Durham  is that the Munich models explicitly follow dark matter haloes even after they are accreted onto larger systems, allowing the dynamics of satellite galaxies residing in the infalling haloes to be followed until the dark matter substructure is  destroyed. The galaxy position is calculated by assigning the galaxy to the most bound particle of a (sub)halo at each time step. This is done until the (sub)halo is no longer identifiable, whereupon the galaxy is assigned to the most bound particle of the (sub)halo at the last time the (sub)halo could be identified. An analytic countdown to galaxy merging begins when the satellite subhalo can no longer be identified, and resets if the parent halo undergoes a major merger. Thus, in the Munich models, the lifetime of galaxies in groups depends on the amount of time the (sub)halo finder can identify the subhalo, plus the analytic countdown.  The analytic merging follows that of \citet{Croton2006}:

\begin{equation}
\label{Eqn:mrg_mun}
 \tau_{mrg} = 1.17 \frac{V_H r^{2}_{sat}}{G m_{sat} \ln \left( 1 +
   \frac{M_H}{M_{sat}} \right)} ,
\end{equation}

\noindent where $\ln \left(1+\frac{M_H}{M_{sat}} \right)$ is the coulomb logarithm, $V_H$ is the halo circular velocity, $r_{sat}$ is the distance of the subhalo from the halo centre at the time it is last identified,  $m_{sat}$ is the mass of the satellite dark halo at the time it was last identified and $M_H$ is the halo mass.

In M\_D06, the amount of reheated (by supernovae) cold gas is proportional to the stellar mass, and the mass ejected from the halo is inversely proportional to the host halo's circular velocity squared \citep{Croton2006}; 

\begin{equation}
     \dot{M}_{reheat} \propto V_{vir}^{-2}\dot{M}_* ,
\end{equation}

\noindent  The important parameters, $V_{vir}$ and $M_*$, have the same relationship in the Durham models. In the Munich models, the reheated mass is proportional to the mass of the halo, while in the Durham model it is proportional to the disk mass. M\_B07 adopts a more sophisticated treatment of supernova feedback. Rather than simply parametrising the effect of supernovae feedback, the M\_B07 model follows the dynamical evolution of the wind as an adiabatic expansion followed by snowploughing. This implementation has the effect of increasing the luminosity of the brightest galaxies. As also occurs with the Durham models, some of the gas will be ejected from low mass haloes in both Munich models.

In the Munich models, a two-mode formalism is adopted for AGN, wherein a high-energy, or ``quasar'' mode occurs subsequent to mergers, and a constant low-energy ``radio'' mode suppresses cooling flows due to the interaction between the gas and the central black hole  \citet{Croton2006}.  In the quasar model, accretion of gas onto the black hole peaks at z $\sim$3, while the radio mode reaches a plateau at z $\sim$2. AGN feedback is assumed to be efficient only in massive haloes, with supernova feedback being more dominant in lower-mass haloes.

The properties of groups in the Munich models, M\_D06 and  M\_B07,  are similar to one another in many cases, as are the properties of the two Durham models, D\_B06 and D\_F08.   Thus, in some analysis, we just discuss the M\_D06 and D\_B06 models, as representatives of their model's lineage. We only discuss the  results based on their descendants, M\_B07 and D\_F08, when they show significantly different behaviour from M\_D06 and D\_B06.

\section{Groups} 

In order to construct a statistically significant (and representative) galaxy group catalogue, we have worked with a set of sub-samples of the Millennium Simulation, amounting to $\sim$3\% of the available volume, -- specifically, 64 boxes of side length 125$h^{-1}$~Mpc drawn from the database.  Our results are robust to the arbitrary selection of the box, having been tested \emph{a posteriori} on alternate boxes of equal size. A luminosity limit of $M_r$=$-$17 in the SDSS $r$-band was imposed. At lower luminosity the effect of the limited mass resolution of the N-body background effects the completeness of the sample. We identify galaxy groups as overdensties in the galaxy population using an FoF algorithm \citep{Geller1983}. No maximum number of members is set but we require that at least four galaxies are linked in order to define a group. Although this removes groups such as the Local group, it follows the \citet{Hickson1982} definition of compact groups.  

We first construct a ``loose group'' (LG) catalogue using a linking length of 0.2 times the mean inter-particle separation,(or in this case, inter-galactic separation). We made this choice by assuming that the galaxies follow the dark matter. This corresponds to a co-moving linking length of $\sim$500$h^{-1}$~kpc. To examine the effects of density we also define two ``compact group'' catalogues, -- Compact (CG), and Very Compact (vCG), Groups -- using co-moving linking lengths of 150$h^{-1}$~kpc and 50$h^{-1}$~kpc, respectively.   The CG linking length of 150$h^{-1}$~kpc is similar to that advised by \citet{McConnachie2008}, based upon their 3D linking length analysis from mock catalogues of Hickson compact groups based on M\_D06. The vCG linking length is comparable to the projected linking length used by \citet{Barton1996} and \citet{Allam2000} to identify groups, a value arrived at by calibrating to the \citet{Hickson1992} catalogue using projected galaxy separations.

We note that the CG galaxies are, by necessity of the group finding algorithm, subsets of the LG catalogue, in that every galaxy assembled into a group at short linking length, must be part of a group with a larger linking length.  Our catalogues also contains clusters and cluster cores, a point to which we return shortly. The physical interpretation of the linking length variation and its impact upon resulting galaxy distribution is non-trivial.  The FoF algorithm essentially probes deeper into the potential well at shorter linking lengths, selecting only galaxies closer to the cluster/group core. These galaxies are generally old, and have sunk deeper into the cluster potential, or they are galaxies near their respective orbital peri-centre.

The algorithm does not only extract the inner region of groups. Galaxies in the outskirts of clusters that happen to be close to one another can also be linked together. This may either be due to a temporary alignment of galaxies that are simply `passing through' that region, or because the galaxies were in close proximity before they entered the cluster, and have not yet been disbursed by tidal forces. These peripheral groups are  essentially cluster substructure, and are a natural part of the analysis. Peripheral groups represent a small fraction of LGs, but represent $\sim$20\% of CGs and vCGs. Also worth noting, the linking length adopted for LGs can include galaxies in haloes beyond the limit of the dark matter halo which the majority of the group members occupy. This means that our LG catalogue is not exactly equivalent to groups that are defined by the halo occupation of galaxies.

\begin{table}
\centering
\caption{The number density of field galaxies, and LG,
  CG and vCG for the D\_B06,M\_D06, M\_B07 and D\_F08 models .}
\label{table_times}
\begin{tabular}{|c|c|c|c|c|}     
\hline 
& D\_B06 & D\_F08 & M\_D06 & M\_B07 \\	
& $h^3$Mpc$^{-3}$ & $h^3$Mpc$^{-3}$ & $h^3$Mpc$^{-3}$ & $h^3$Mpc$^{-3}$ \\
\hline									
Field &	 $6.5\times 10^{2}$ &	$6.6\times 10^{2}$ & $6.3\times 10^{2}$ &	$5.0\times 10^{2}$   	\\
LG    &	 $2.1\times 10^{3}$ &	$2.2\times 10^{3}$ & $2.0\times 10^{3}$ &	$1.6\times 10^{3}$   	\\
CG    &	 $1.3\times 10^{3}$ &	$1.4\times 10^{3}$ & $8.9\times 10^{4}$ &	$8.7\times 10^{4}$   	\\
vCG   &	 $5.4\times 10^{4}$ &	$6.6\times 10^{4}$ & $1.2\times 10^{4}$ &	$1.4\times 10^{4}$   	\\
\hline
\end{tabular}
\end{table}

\subsection{Merging Timescales}
We compare the merging histories and timescales of haloes and galaxies within the models. Any differences will be important in determining the origin of various group properties.  To compare  the lifetime of satellites we look at a) the lifetimes of satellites that have merged, and thus contribute to the mass and Luminosity of the Central Galaxy, and b) the lifetime of satellites that have not yet merged, and are hence satellites at $z=0$, and contribute to the Group catalogues.

\begin{figure}
\includegraphics[width=84mm]{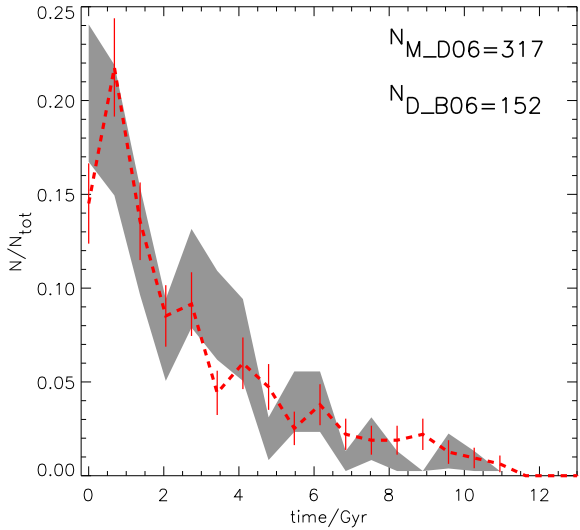}
\caption{Shows the distribution of galaxy merging times. The y-axis is number of galaxies in each bin normalised by the total number of galaxy mergers. The black line is the D\_B06 model and the red dotted line is for the M\_D06 model. Errors are Poisson.  The number of merging galaxies in each model is quoted in the panel. }
\label{Fig:mtime}
\end{figure}

Figure~\ref{Fig:mtime} shows the distribution of merging times, defined as the time between a satellite first entering a halo and being totally merged with the central galaxy. We select the same 19 millennium simulation clusters from each model to make a fair comparison between the codes and do not put any limit on the luminosity of infalling satellites. The distribution of merging times shows little difference between the models.  The average M\_D06 satellite has lasted 5.8 Gyr, while D\_B06 satellites last 6.2 Gyr, with standard deviations of $\sim 1$. We note, however, that twice as many satellites have merged in the Munich model (317) than in the Durham model (152). 

\begin{figure}
\includegraphics[width=84mm]{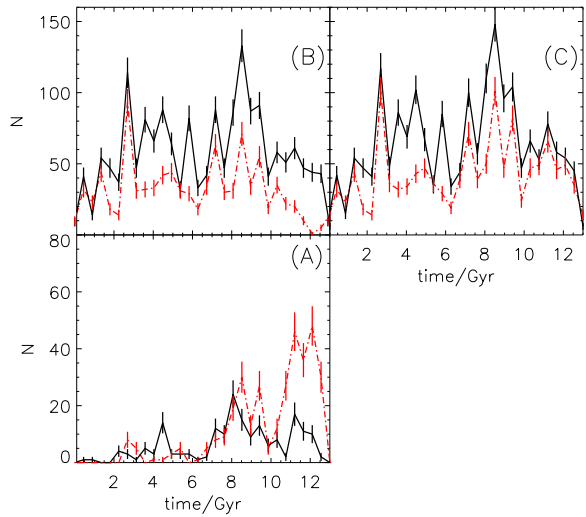}
\caption{{\bf Panel A:} Shows the number of galaxies which merge with the halo central galaxy during the simulation against the time the galaxy entered the host halo. {\bf Panel B:} Shows the infall time of those galaxies which are still satellites at z=0. {\bf Panel C:} is the sum of the plots in Panel A and Panel B. The solid black line shows the D\_B06 result and the red dashed line is for M\_D06.}
\label{Fig:mtime2}
\end{figure}

This result seems a little contradictory, as similar merging timescales should result in similar numbers of mergers. We look to the number of satellites which do not merge in order to reconcile this. Figure \ref{Fig:mtime2} shows the time of infall for the satellites into the host halo, for satellites that merge (Panel A), satellites that do not merge but rather remain as satellites at z=0 (Panel B), and all satellites (Panel C). Again, the fact that more satellites  merge in M\_D06 compared with D\_B06, can be seen in panel A, but it can be seen that the difference is dominated by satellites which fall in to the host halo at early times. By contrast, the number of satellites which do not merge at all is significantly larger for D\_B06, compared with M\_D06. The difference is greatest for satellites which accrete early. This significant population of satellites which do not merge by z=0 can be expected to affect the properties of groups which we present in the remainder of this study, both because of the lower number of satellites ``feeding" the central galaxy in D\_B06, and also the larger numbers of satellites which survive to be included in the group catalogues. Panel C shows the infall time of all satellites from our sample of 19 clusters, regardless of whether they merged,  and shows that the differences between the models being relatively minor. 

There are more haloes falling into the Durham models and fewer galaxy mergers. The galaxies \emph{which merge} have the same lifetime in each model. If we look at the time of infall of galaxies which merge, and those that remain in the cluster at z=0, (Fig. \ref{Fig:mtime2}), there is a substantial population of galaxies which never merge. In the Durham models more galaxies survive than in the Munich models (panel (B) of Fig.2). Further, we also see that more (of the early accreted) galaxies merge with the central object in the Munich model (panel (A) of Fig.2).At the earliest times in the M\_D06 model almost all the galaxies have merged, while in the D\_B06 model they have not. Thus, the `maximum' lifetime of satellite galaxies is $10$Gyr in M\_D06 but $>13$Gyr in D\_B06. 

 While not shown, our examination of the merger trees also found that in the D\_B06 model a greater number of haloes merge with the 19 clusters. This is because subhaloes are not followed when constructing the merger trees in the Durham models. In the Munich models there is a delay in the halo merger time relative to the Durham models became infalling haloes are able to enter a subhalo stage before they are considered merged. When the (sub)halo in each model is no longer identifiable, the SAMs are used to calculate the lifetimes of galaxies. In the Durham models there appears to be an extremally long lived population of satellites which are not present in the Munich

\section{Observations}

 We use the observational loose group catalogues of \citet{Tago2008}, \citet{Yang2008}, and \citet{Tucker2000},  and the \citet{Allam2000} compact group catalogue. These group catalogues take large redshift surveys of galaxies and use a FoF algorithm to assemble them into groups. \citeauthor{Tago2008} (\citeyear{Tago2008}, SDSS5\_T08) implemented a standard FoF algorithm which scales according to the distance, and applied it to the SDSS DR5 \citep{Adelman2007}. The initial linking length is 0.25$h^{-1}$Mpc in projection, and $250kms^{-1}$ along the line of sight. \citeauthor{Tucker2000} (\citeyear{Tucker2000}, LCRS\_T00 ) catalogue uses the Las Campanas Redshift Survey \citep{Shectman1996}, and applies a fiducial linking length of 0.715$h^{-1}$Mpc and $500kms^{-1}$, scaled from this value at $z=0.1$ and rising to 1.8 times this at $z=1.7$. \citeauthor{Allam2000} (\citeyear{Allam2000}, LCRS\_A00) uses the same catalogue as LCRS\_T00 but a shorter linking length of 0.05$h^{-1}$Mpc and $500kms^{-1}$. Finally, \citeauthor{Yang2008} (\citeyear{Yang2008}, SDSS4\_Y08) uses a more complicated iterative approach, which, nevertheless, includes a FoF algorithm at its core \citep{Yang2005}. This algorithm applied to SDSS DR4 \citep{Adelman2006}.

Our synthetic group catalogues are compared directly with these empirical datasets wherever such comparisons are possible and meaningful within our analysis, bearing in mind limitations in the models, and in the observations. Where appropriate, ``cuts" are made in the synthetic catalogues to mimic observations.

\section{Results}
\subsection{Density of Groups in the Models}  
\label{Sec:den}
The difference in the proportion of galaxies in groups should provide a key diagnostic for comparing and discriminating between the four SAMs, relative to empirical data. 
Within the Millennium Simulation volume at z=0, the four SAMs under consideration yield differing numbers of galaxies, despite being built upon the same underlying dark matter distribution.  It is useful to review the respective galaxy numbers and distributions for these models. Restricting the discussion to the relevant sampling criteria (i.e., M$_r$$<$$-$17) the  M\_D06 and  M\_B07, and D\_B06 and D\_F08 models have a galaxy number density given in Table \ref{table_times}. The number density of galaxies in the field for each model is quite similar, as well as in LGs. However, the Durham models have significantly denser galaxy populations within  CGs and vCGs. The physics which the different models employ to account for group environments appear to have significantly altered the  nature of   compact groups.

The relative proportions  galaxies that are classified as being members of groups, along with the average group richness, are listed in Table~\ref{numbers}. In this instance we define group richness as simply being the number of galaxies in a group.

The percentage of galaxies associated with LGs (and the numbers of galaxies per loose group) is comparable between three of the four SAM variants, with the M\_D06 model showing  approximately 6 percent fewer groups than the D\_B06 model. The  models  diverge increasingly with decreasing linking length, with the Munich models, M\_D06 and M\_B07,  having 5 to 6  times fewer vCGs than in the Durham models.  The M\_D06 model produces noticeably fewer rich groups at all linking lengths compared with the other Munich model, M\_B07. This indicates that the implementation of supernovae has a significant effect on the richness of group catalogues. The Durham models,  D\_B06 and D\_F08, have slightly richer loose groups than the Munich models. This is associated with the smaller populations of medium brightness red galaxies in the Munich models, which may be the result of the differences in the creation of halo catalogues, the tracing of subhalo mergers, or due to radio mode AGN feedback.

\begin{table}
\label{numbers}
\centering
\caption{Top: Percentage of galaxies in FoF groups for LG, CG and vCG
  in each of the four SAMs; Bottom: Mean group richness for groups in
  the four models}
\begin{tabular}{|c|c|c|c|}     
\hline   
          & LG & CG & vCG \\
\hline
Bower (D\_B06)     & 44 & 18 & 5.5 \\
Font (D\_F08)      & 44 & 19 & 6.0 \\
De Lucia (M\_D06)  & 38 & 10 & 0.9 \\
Bertone (M\_B07)   & 43 & 14 & 1.3 \\
\hline
Bower (D\_B06)     & 13.5 & 9.4 & 6.6 \\
Font (D\_F08)      & 13.1 & 9.2 & 6.6 \\
De Lucia (M\_D06)  & 11.5 & 7.5 & 4.7 \\
Bertone (M\_B07)   & 13.6 & 8.0 & 5.0 \\
\hline
\end{tabular}
\end{table}

\citet{McConnachie2008, McConnachie2009} compared compact groups in mock redshift catalogues to SDSS DR6 observations, and concluded that the M\_D06 SAM overproduces CGs by $\sim$50\%. By extension, Table~\ref{numbers} indicates that the D\_B06 and D\_F08 models result in an even more dramatic ``overproduction'' of CGs (by an order-of-magnitude).  Thus, in this regime, none of the models are good fits to the empirical data, and  the Durham models are particularly poor. 

\begin{figure}
\includegraphics[width=80mm]{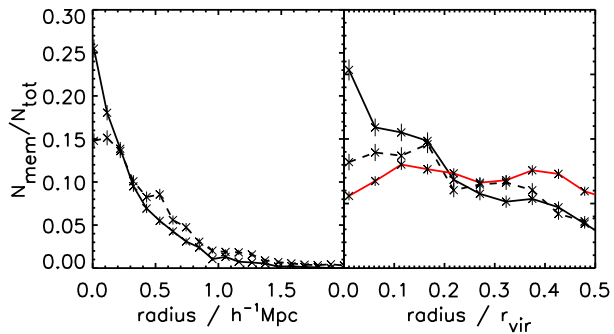}
\caption{{\bf Left Panel:} The number of galaxies in LGs against distance from the centre, normalised by the total number of galaxies.  The solid line is for D\_B06 LGs and the dashed line is for M\_D06 LGs. {\bf Right Panel:} The normalised number of galaxies against the radius scaled by the virial radius of the host.The red line shows SDSS4\_Y08 LGs for the 60 most massive clusters. Errors are Poisson. There are 2294, 2022 and 4688 in the D\_B06, M\_D06 and SDSS4\_Y08 datasets respectively.}
\label{Fig:galpos}
\end{figure}

\begin{figure}
\centering
\includegraphics[width=84mm,clip]{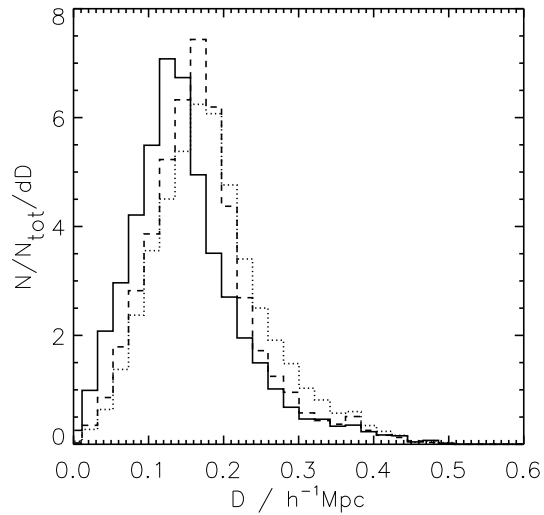}
\caption{The distribution of nearest neighbour pair separations for galaxies in LGs for the three models, the solid line is D\_B06,  the   dashed line is M\_D06, and the dotted line is D\_B07. The plot shows that D\_B06 galaxies tend to be closer together on average. } 
\label{nearest}
\end{figure}

 Figure \ref{Fig:galpos} shows the  galaxy distribution of galaxies with radius for the LGs of M\_D06 and D\_B06 models.   The left panel shows the number of satellites versus radius in Mpc.  The right panel of  Fig. \ref{Fig:galpos} shows the  same plot with the radius scaled by the virial radius. Both panels show that there are more  satellites in the inner regions  of the  D\_B06 LGs, relative to the outer regions, compared with the M\_D06 model. The red line in the right panel shows the 60 most massive SDSS4\_Y08 observed Loose Groups, making them similar in mass range to the population used in making the simulation plots. This indicates that both models have a radial distribution of satellites which is more centrally concentrated than the observations, with the problem being particularly acute in the D\_B06 model. 

This more concentrated distribution of satellites is also reflected in the distribution of nearest neighbour pair separation distances for LG members, seen in   Fig.~\ref{nearest}.  It can be seen that  the D\_B06 model the nearest neighbour separation is  smaller than for the M\_D06 and M\_B07. We note that the D\_F08 model groups have very similar properties to D\_B06 and is thus not included in these figures.

\subsection{Luminosity Functions}

\begin{figure}
\centering
\includegraphics[width=84mm,clip]{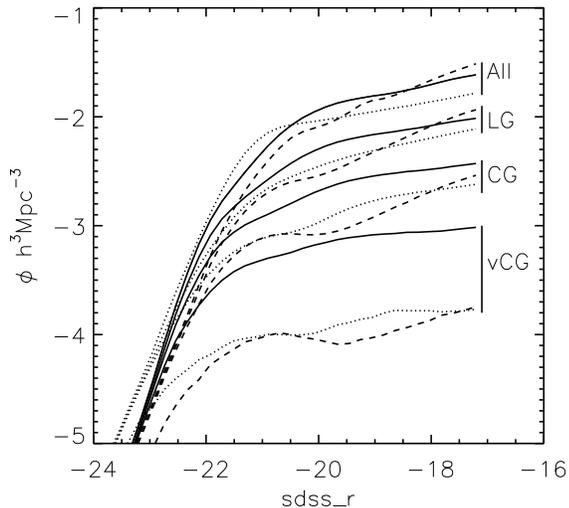}
\caption{Derived SDSS r-band luminosity functions for galaxy
groups constructed with the SAMs described herein:
``All'' refers to the global galaxy luminosity function; ``LG'' to
loose groups; ``CG'' to compact groups; ``vCG'' to very compact
groups. The solid/dashed/dotted lines refer to the D\_B06/M\_D06/M\_B07
models, respectively.}  
\label{fig:AllLum}
\end{figure}

We now plot the group luminosity functions. Figure \ref{fig:AllLum} shows how the luminosity function changes in the different group environments for  the SAM models. It is unsurprising that the global luminosity functions of all the models are similar, because the semi-analytic models are designed to replicate the same observational luminosity function of \citet{Blanton2003}.  At smaller linking lengths (moving from top to bottom in the figure) there is a decreasing number of galaxies in all SAMS, a fact which is more dramatic in the Munich variants, as noted in \S~\ref{Sec:den}. This provides further evidence that the Durham model galaxies are more centrally concentrated than those of the Munich variants. 

The second feature of Fig. \ref{fig:AllLum} is the relative dearth of intermediate luminosity ($-$21$\simlt$M$_r$$\simlt$$-$18) galaxies in the M\_D06 catalogues (in relation to a simple Schechter (1976) function).  This is manifest in the ``wiggle'' or ``dip'' seen in the M\_D06 group luminosity function. This feature is not present in the other models. This wiggle becomes more apparent at shorter linking lengths, (i.e., CGs and vCGs). The M\_B07 SAM, which employs the same AGN feedback prescription as M\_D06, shows no such feature in the luminosity function. Weinmann et~al. (2006; fig~3) show a similar wiggle in the luminosity function of groups in particular mass bins, i.e. the conditional luminosity function, using the Croton et~al. (2006) SAM, which is a close ``cousin'' to the M\_D06 SAM employed here. 

\begin{figure*}
\centering
\includegraphics[width=120mm,clip]{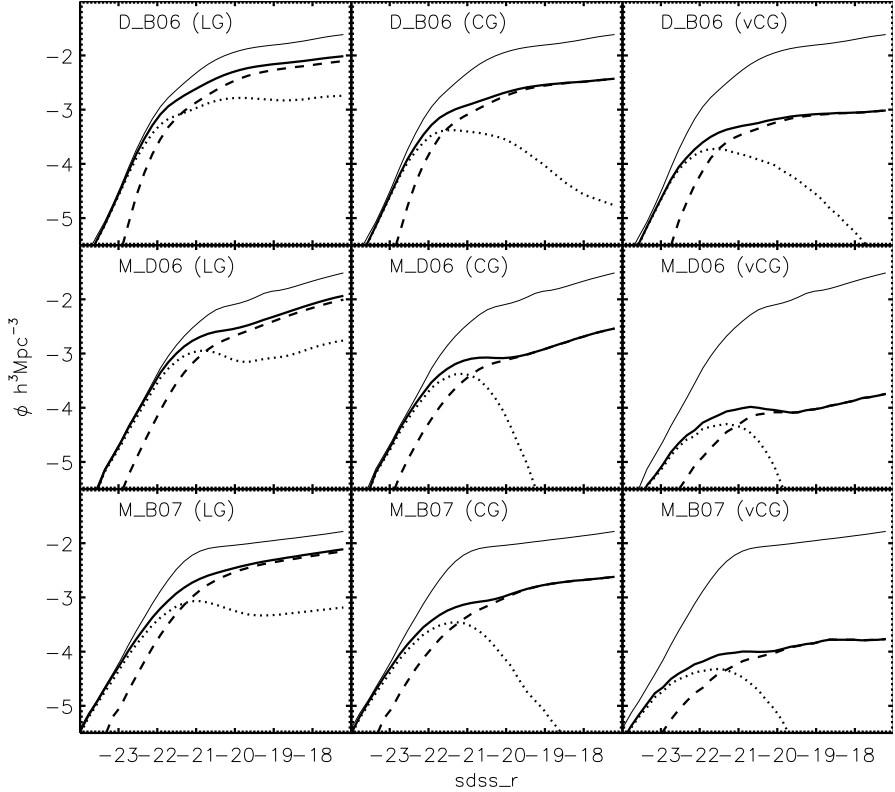}
\caption{Group luminosity functions for the three primary
SAMs under consideration (D\_B06, M\_D06, M\_B07). In each panel, the 
dotted lines correspond to central galaxies and the dashed lines are
for the satellites; the thick solid lines represent the group luminosity
function and the thin solid line defines the global galaxy luminosity
function for reference. The legend D\_B06, M\_D06, M\_B07 refers to the model
as defined above and the number in brackets refers to the linking length in 
h$^{-1}$kpc, so the first column is for LGs, the centre column is for CGs and
the right column is for vCGs.}
\label{fig:LFsatcen}
\end{figure*}

In order to better understand the origin of the shape of these luminosity functions, we decompose 
the luminosity functions of the three primary SAMs (D\_B06, M\_D06, M\_B07) and the three primary linking lengths (LG, CG, vCG) under consideration here, into central galaxies and satellites (Fig.~\ref{fig:LFsatcen}). The global galaxy luminosity function is also shown for reference. The M\_D06 (middle row), and M\_B07 (bottom row) centrals show a Gaussian distribution, while the D\_B06 model, (top row),  centrals are better described by a Schechter function. There is an increase in the number of faint central galaxies in LGs compared with CGs and vCGs, which alters the shape of the central luminosity function as we move to larger linking lengths. All satellites are distributed roughly according to  Schechter functions. 

 The wiggle in the M\_D06 luminosity function is  due to the combined effect of (i) a general lack of satellites, and (ii) the "peaked" nature of the central galaxy luminosity distribution.  Relative to the D\_B06 SAM prediction, this distribution of centrals is very narrow, without the low-luminosity tail associated with the D\_B06 model.  This effect is particularly apparent at  the shortest linking lengths, i.e. in vCGs. The greater total number of vCGs in the Durham models compared to the Munich models is also clear in the vCG panels, again reflecting their cental concentration.  

The contrast between the M\_D06 and M\_B07 models is of particular interest because they use the same AGN feedback implementation, but differ in their choice of supernovae feedback. The more sophisticated, and more effective SNe model of M\_B07 makes the luminosity distribution of centrals wider by making a tail towards fainter magnitudes that is not present in M\_D06. The most luminous galaxies are also more luminous in M\_B07 than in M\_D06. The M\_B07 treatment leads to a more significant population of intermediate luminosity satellites with a shallower luminosity function, thus smoothing the wiggle. A lower number of low-luminosity satellite galaxies in such a shallow luminosity function is accompanied by ``over-luminous'' massive (luminous) galaxies  \citep{Bertone2007}.  M\_B07 shows a decrease in the number of low luminosity satellites and an increase in the number of intermediate luminosity galaxies, reflecting a SN  feedback in M\_B07 which is stronger in dwarfs and weaker in large haloes. The increase in high luminosity galaxies and decrease in low luminosity galaxies is noted in \citet{Bertone2007} which they suggest could be solved by increasing the supernova feedback time or increasing the effect of AGN feedback. It also produces a wider central galaxy distribution by not only increasing the number of very bright galaxies but also the population of dim centrals.
  
\subsection{Brightest Group Galaxy}

We define three types of identified groups. The first type, bright central groups, are those where the brightest galaxy is also the central. The second type, peripheral groups, are those without a central galaxy and, therefore, the brightest galaxy is a satellite. The third type of group, dim centrals, are those where the central galaxy is not the brightest. Except in the D\_F08 model, the  central galaxy is the only one with hot gas and it acquires all the hot gas from infalling satellites. The central galaxy is also the only  galaxy that experiences mergers and grows hierarchically. Table 3 shows the populations of these three types for LG, CG and vCG, the format being, groups with bright central groups/peripheral groups/dim central groups. 

\begin{table}
\caption{Proportion of galaxy group types where the first number is
  the proportion of bright central, peripheral and dim central groups.}

\centering
\label{table_type}\begin{tabular}{|c|c|c|c|}    
\hline   
Model  & LG  & CG & vCG  \\
\hline
D\_B06    & 79/00/21 & 63/12/25 & 64/15/21 \\
M\_D06 & 90/01/09 & 74/21/05 & 75/20/05  \\ 
M\_B07  & 82/02/16 & 70/22/08 & 71/23/06   \\

\hline
\end{tabular}
\end{table}

There is a  difference between the three main models for LGs. The Durham models have a smaller fraction of Bright Central Groups relative to dim centrals,   compared with the Munich models. This discrepancy continues to smaller linking lengths. This likely driven by the greater number of mergers in the Munich models, feeding the growth of the central galaxy. The more sophisticated supernova feedback of the M\_B07 model has decreased the fraction of BGGs. The Munich models also have more peripheral groups at all linking lengths      
                    
\begin{figure*}
\centering
\includegraphics[width=120mm,clip]{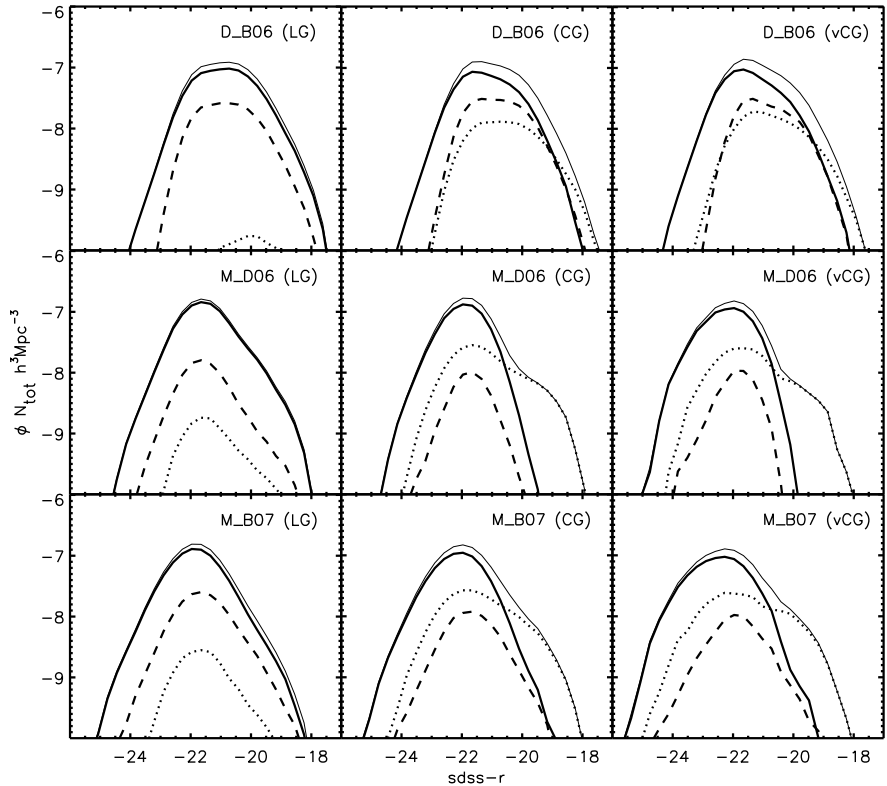}
\caption{Luminosity function of all first ranked galaxies (the brightest galaxy in each group, thin solid line); distribution of first ranked galaxies in bright central groups (thick solid line); the first ranked galaxies in peripheral groups (dotted line)  and the first ranked galaxies in dim central groups (dashed line). }
\label{fig:LFtype}
\end{figure*}

In Fig. \ref{fig:LFtype} the luminosity functions of the first ranked galaxies (i.e. the brightest galaxy in each group) of our groups have been plotted. These have then been decomposed by group type, with distribution of first ranked galaxies in bright central groups, the first ranked galaxies in peripheral groups  and the first ranked galaxies in dim central groups. It can be seen that, in the Munich SAMs, for the denser groups,  there is a large difference between the shape of the LF of the brightest galaxies in  central and peripheral groups. The difference is most extreme for the M\_D06 model, where the low magnitude tail is due, almost entirely, to peripheral groups. On the contrary, in the D\_B06 model the distribution of groups is not particularly different for the different group types.

\subsection{Halo-based Groups}
In this section groups consist of galaxies which lie within the same dark  matter halo, whose extent is determined by using a density contour defined by a dark
matter particle separation of 0.2 times the mean inter-particle separation. This is not to be confused with the linking length used to defined groups which acts on
galaxies rather than dark matter particles. Groups determined in this manner differ from those determined by using FoF algorithms. Even for LGs there is not a one-to-one correspondence between halo groups and FoF groups. The limit on the minimum number of galaxies used to define a group remains at four. 

\begin{figure*}
\centering
\includegraphics[width=120mm,clip]{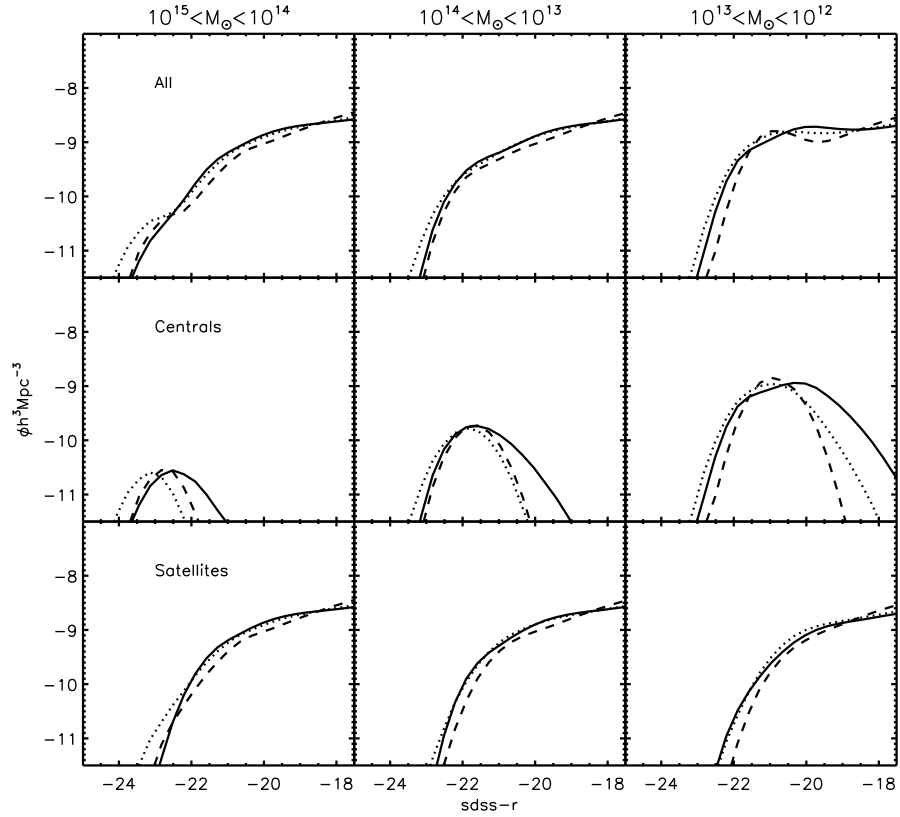}
\caption{The conditional luminosity functions in three mass bins which correspond roughly to clusters, large groups and small groups, for halo-based groups for the three primary SAMs employed here, at redshift $z$=0. {\bf The top row} shows the distributions of all halo members, {\bf the second row} show central galaxies and {\bf the bottom row} shows satellites.  The solid, dashed and dotted lines are for the D\_B06, M\_D06 and M\_B07 models respectively.}
\label{fig:CLF}
\end{figure*}

We examine the conditional luminosity functions in three different group mass bins. These conditional luminosity functions are plotted in Fig.~\ref{fig:CLF}, for the D\_B06, M\_D06, and M\_B07 SAMs. The luminosity functions are further separated into centrals (middle row) and satellites (bottom row).

The $10^{13-14}M_\odot$ and $10^{14-15}M_\odot$ mass bins show little evidence for the ``wiggle'' (for any of the SAMs). The ``wiggle'' in the M\_D06 model, and to a lesser extent the M\_B07 model, is particularly prominent in the lowest mass bin (top right panel), where a number of physical processes become relevant. Bower et al. (2006) points out that at $\sim 2\times10^{11}M_{\odot}$ the cooling rate exceeds the free-fall rate and the halo is no longer in hydrostatic equilibrium. This has repercussions for the effectiveness of feedback from the central source \citep{Binney2004} and is used in the Durham paper to explain the break in the luminosity function. Our results indicate that in the Munich models the processes occurring in the mass range may have other repercussions.

For the highest mass bin we can see that the satellite luminosity function is steepest for the M\_D06 SAM and shallowest for the D\_B06 model. The characteristic luminosity at the ``knee'' of the Schechter function ($M_*$) is lowest for the D\_B06 model. However, this bin has only a small effect on the ``wiggle'' alluded to in Section 5.2, which occurs at lower luminosity than the wiggle seen in this particular mass bin.  In the $10^{12-13}M_\odot$ bin, where the wiggle is prominent in the M\_D06 model, the satellite distribution is fairly steep and the central galaxy luminosity function relatively narrow and bright.  By contrast the D\_B06 model has a  much broader central galaxy luminosity function, indicating a  tendency to produce significantly more low-luminosity centrals compared to M\_D06. Again, the unmerged satellites in the Durham models mean less feeding of the central galaxy resulting in this tail to lower luminosities.

\begin{figure*}
\centering
\subfigure[Satellite galaxies]{
\includegraphics[scale=0.4]{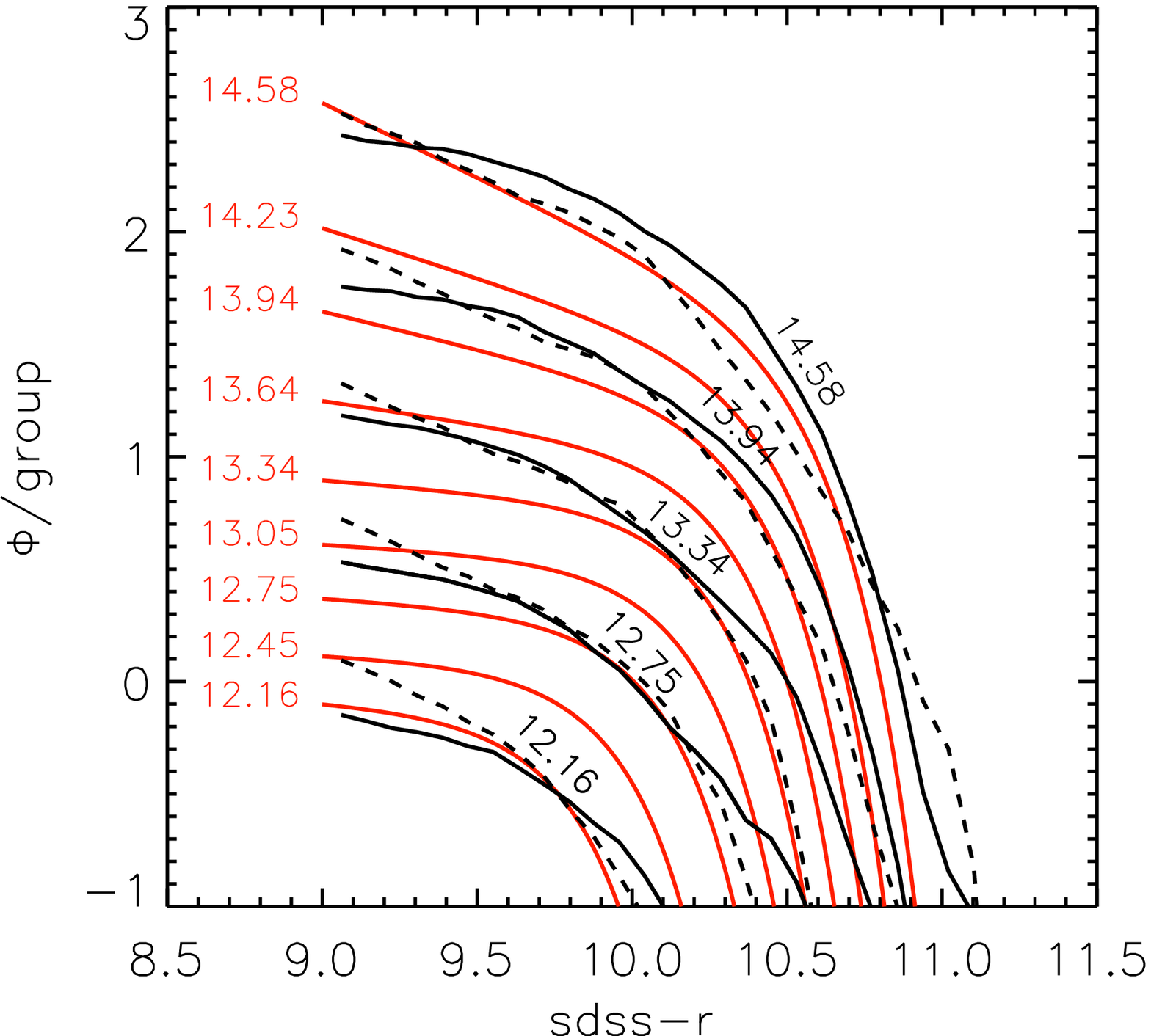}
\label{fig:satlumclf}
}
\subfigure[Central Galaxies]{
\includegraphics[scale=0.4]{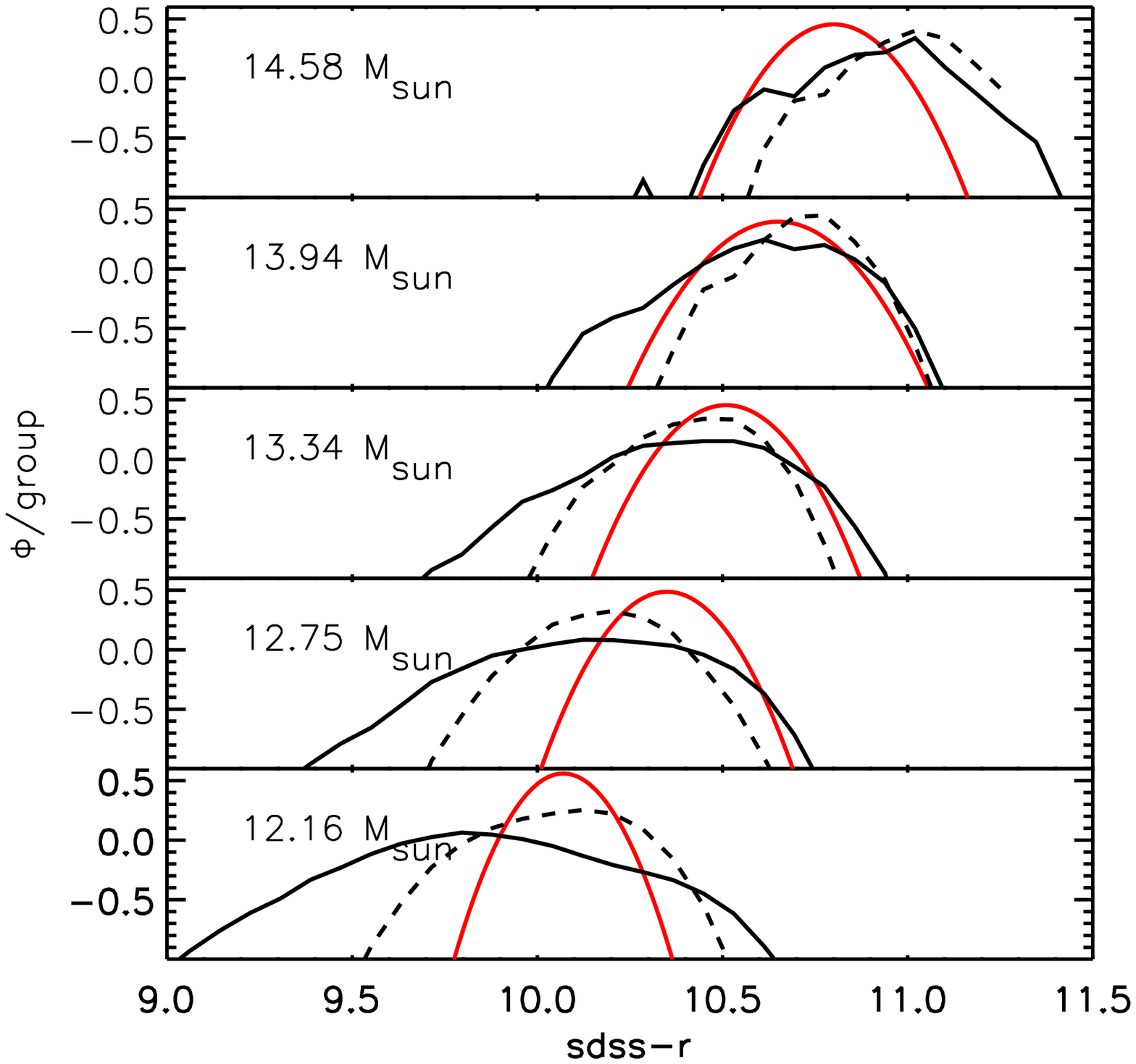}
\label{fig:cenlumclf}
} 

\label{fig:obsCLF}
\caption[Optional caption for list of figures]{The conditional luminosity function of the models, D\_B06 (solid black line) and M\_D06 (dashed line) plotted over the CLFs of SDSS4\_Y08 (shown in red). The number quoted being the centre of each mass bin and the box width being $\sim0.31$ mag. In panel `a' the top pair of lines are the $10^{14.58}M_\odot$ cut, the next is the  $10^{13.94}M_\odot$ cut etc, down to $10^{12.16}M_\odot$  }
\end{figure*}

In order to compare this with observations we over plot the results of M\_D06 and D\_B06 onto the conditional luminosity functions provided by SDSS4\_Y08. \citet{Weinmann2006} note that the method used by SDSS4\_Y08, presented in Yang et al. (2005), artificially narrows the central galaxy luminosity function. This is because their iterative technique uses the brightest galaxy luminosity in the derivation of the group halo mass, while in the models there is no such direct linking of mass and luminosity. However, the difference is not enough to affect our comparison. SDSS4\_Y08 shows the CLFs for groups using SDSS DR4 galaxies, and are  fit with modified Schechter and Gaussian functions to the satellite and central galaxy luminosity distributions respectively. The functional forms of the fits are,

\begin{equation}
\label{eqn:central}
\Phi_{cen}(L|M) = \frac{1}{\sqrt{2 \pi}\sigma_c}
              \exp\left[\frac{(\log{L}-\log{L_c})^2}{2\sigma_c^2} \right],
\end{equation}

\begin{equation}
\label{eqn:satellite}
\Phi_{sat}(L|M) = \phi^*_s \left( \frac{L}{L^*_s} \right)^{(\alpha^*_s+1)} 
       \exp \left[-\left(\frac{L}{L^*_s} \right)^2 \right],
\end{equation}

\noindent where, L is the luminosity, $L_c$ is the mean position of the Gaussian, $\sigma_c$ is the width of the Gaussian, $\phi^*_s$ is the normalisation of the modified Schechter function, $\alpha^*_s$ is the low mass slope and $\log(L^*_s) = \log(L^*_c)-0.25$ and is the position of the knee of the modified Schechter function. 

SDSS4\_Y08 provide the best fit parameters, which we compare to the M\_D06 and D\_B06 models using the same mass bins as SDSS4\_Y08 in Fig. \ref{fig:satlumclf} and Fig. \ref{fig:cenlumclf}, where Panel `a' shows the satellite galaxy distribution and Panel 'b' the central galaxy distribution. The highest mass bin in the models corresponds most closely with the observations.

 The shape of the conditional luminosity functions for the satellites are considerably different for the CLFs. At all masses, there are far fewer low luminosity satellites in the models than than in the observations. The discrepancy is less severe for the  M\_D06 model for high mass clusters. The difference between observations and models for the satellite luminosity function is larger as we go to lower mass. As we go to $10^{12.75}$ M$_\odot$ and below, it is the  D\_B07 groups which have more low luminosity satellite galaxies.

The central galaxy conditional luminosity functions also differ significantly between models and observations.  The M\_D06 model shows that the low mass ($10^{12.16}$ M$_\odot$)  group centrals peak in the same place as the observations, while the mass bins are somewhat displaced, both toward lower luminosities ($10^{12.75}$ M$_\odot$ and $10^{13.3}$ M$_\odot$) and higher luminosities ($10^{13.94}$ M$_\odot$ and $10^{14.58}$ M$_\odot$). The M\_D06 model  is also broader than the observations. These discrepancies are even greater for the D\_B06 groups, which even wider than the M\_D06 groups.

 The median vCG has a mass of $1.7\times 10^{13}M_\odot$ for both M\_D06 and D\_B06, but with considerable variation. This means that the distributions closest to this value are more important to the analysis of vCGs than further away. In Fig. \ref{fig:satlumclf} and \ref{fig:cenlumclf} we see that the models are reasonably similar at this point, although the peak in the central luminosity function is greater in the Munich variant.  This  suggests that the significant lack of vCGs in M\_D06 compared to D\_B06 is due to the positions  of galaxies in groups rather than the absolute numbers, and we refer the reader back to Figure~3. 

\section{Magnitude Gap}

\begin{figure*}
\centering
\includegraphics[trim = 0mm 0mm 0mm 0mm,clip, width=160mm]
{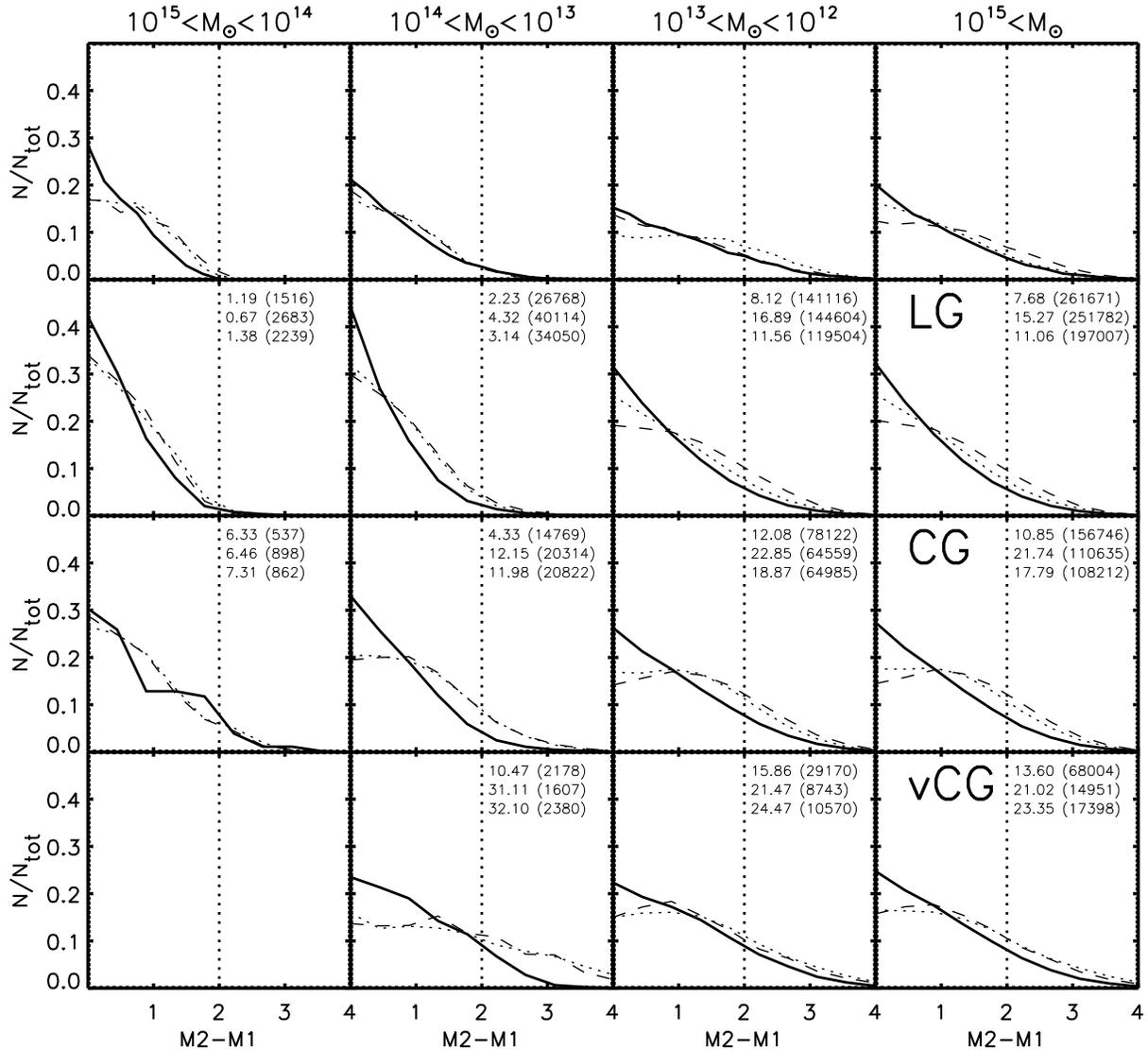}
\caption{Conditional (mass-dependent) magnitude gaps between first- and second-ranked galaxies for the SAMs included in this study. Solid, dashed and dotted lines are for D\_B06, M\_D06 and M\_B07 groups respectively. {\bf The top row} is for halo-based groups divided by group mass and {\bf the far right column} is for all FoF-groups regardless of mass, specifically, {\bf row 2} is for LGs, {\bf row 3} is for CGs and {\bf row 4} is for vCGs. The subsequent panels show the FoF based groups broken down by halo mass. The vertical line shows the cut-off for Fossil groups (Sales et al. 2007). The numbers in each panel give the percentage of groups which are fossil systems, and, in brackets, the total number of groups in each mass bin. The first number is for D\_B06, the second for M\_D06 and the third for M\_B07. } 
\label{fig:offset}
\end{figure*}

The magnitude ``gap'' between the first- and second-ranked (and, indeed, lower-ranked) galaxies within a group can be used as a  predictor of  group (or halo) age \citep{vonBendaBeckmann2008}, as the central galaxy tends to grow unceasingly with time via satellite accretion/stripping. This mechanism inevitably increases the magnitude gap. This process is controlled by feedback in the central galaxy (\S 2) and by infalling galaxies. Taken to its extreme, such an effect gives rise to the so-called ``fossil groups'', which are groups with a magnitude gap greater than two, most likely caused by a lack of recent galaxy infall onto the group (\citealt{DOnghia2005, SommerLarsen2006, Dariush2007, Sales2007, vandenBosch2007, DiazGimenez2008, MendesdeOliveira2006, Vikhlinin1999}, however, see \citealt{Zibetti2009}). 

In Fig. \ref{fig:offset}, far right column, we compare the magnitude gap distribution between first and second ranked group galaxies in the suite of SAMs employed here. We show the distribution for all groups (top panel) as well as for the LGs, CGs and vCGs separately as we move downwards.   The M\_D06 model (and to a lesser extent, that of M\_B07) shows a preferred magnitude gap of $\sim$1~mag between the two most-luminous galaxies in the model groups, (particularly in the CGs and vCGs), while the D\_B06 SAM predicts far more equal luminosity first- and second-ranked group galaxies. The difference in the distributions between the models is quite apparent, with a ``turn over'' in the two Munich models, i.e. both are ``flatter''  and ``broader'' than the Durham models,  for all group types and in all mass bins.  Again, a significant difference in how the model's  galaxies evolve within dense environments has been highlighted by these observable characteristics.  

Dariush et~al. (2007; fig~4a) show a comparable representation of the top-left panel of our Fig. \ref{fig:offset}, employing the \citet{Croton2006} SAM as applied to the Millennium Simulation (in the $R$-band), and for a slightly different mass range, but effectively similar to what we have shown.  Dariush et al. point out that the magnitude gap distribution of LGs in the \citet{Croton2006} model is similar to the ln~$\Lambda$=2 theoretical model of \citet{Milosavljevic2006}, where ln~$\Lambda$ is the Coulomb logarithm that controls the merger rate. When the \citet{Croton2006} SAM is compared with the SDSS C4 catalogue \citep{Miller2005}, as is done in fig.~4c of Dariush et~al,. the apparent mismatch at small first- and second-ranked magnitude differences between the Munich SAM, and the data become apparent - i.e., the SDSS C4 catalogue shows a magnitude difference distribution which prefers approximately equal luminosity first- and second-ranked galaxies in groups and clusters, more consistent with the Durham SAM predictions.

If we compare our result to the halo occupation distributions of \citet{vandenBosch2007}, we find that we have far fewer fossils groups in the two high mass bins for all three models but more for M\_D06 groups in the lowest mass bin. A halo occupation distribution is a statistical model of the number and luminosity of galaxies occupying a dark matter halo of a given mass. As such, it is strongly related to the conditional luminosity function previously discussed, and  serve as a base of comparison for the SAMs.

\begin{figure*}
\centering
\includegraphics[trim = 0mm 0mm 0mm 0mm,clip,
  width=120mm]{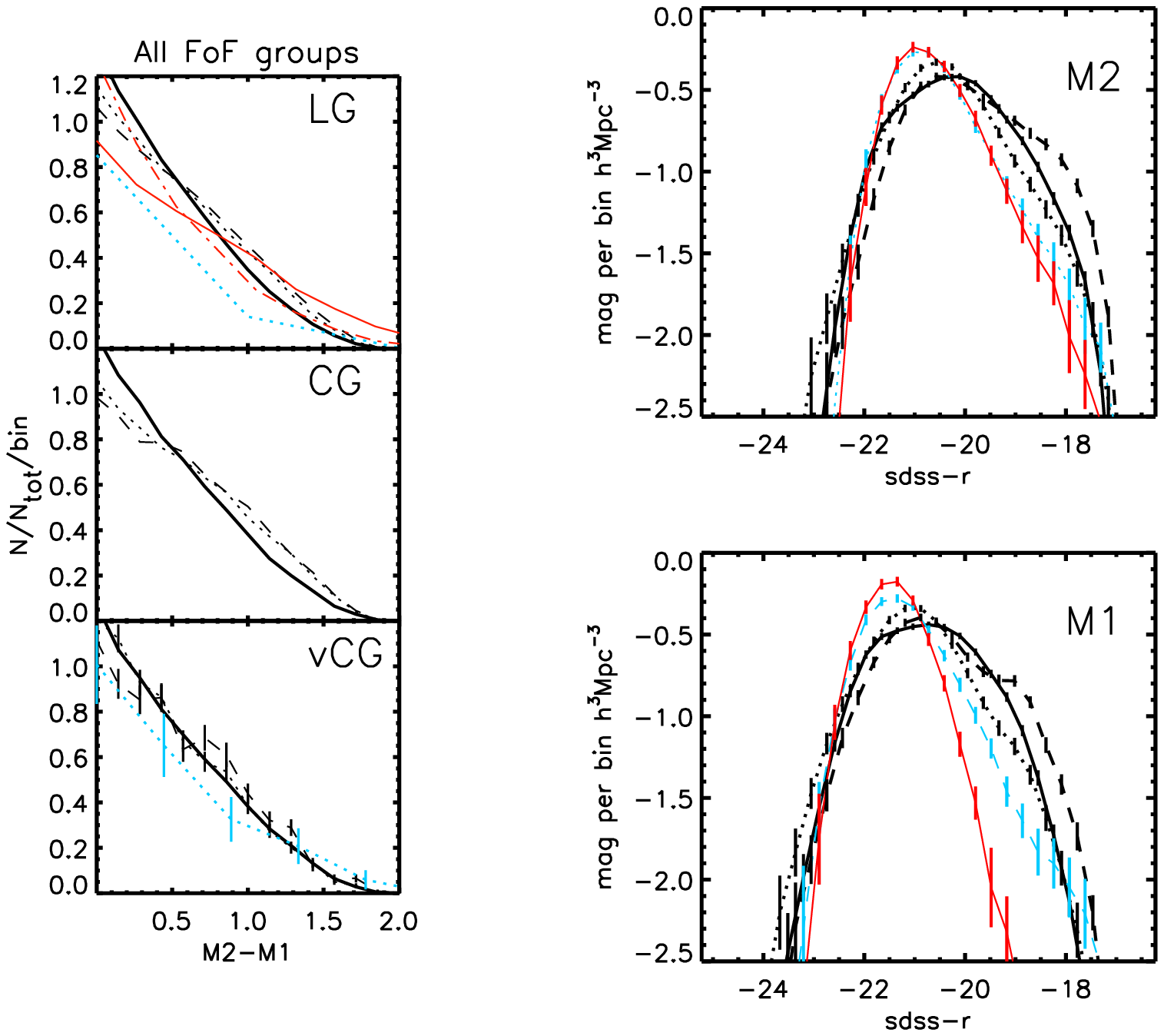} 
\caption{{\bf Left column:}  Distribution of the magnitude gap between 1st and 2nd ranked galaxies, normalised by total number of groups in each sample. The thick solid, dashed and dotted lines are for D\_B06, M\_D06, M\_B07 models. The coloured lines show observational data. The top left panel shows Tago et al. (2008) groups (red dashed line), SDSS4\_Y08 (red solid line) and Tucker et al. 2000 LGs (blue dotted line), the middle panel shows the results for CGs and and the bottom left panel shows Allam et al. (2000) compact groups (blue dotted line) along with vCGs. Poisson counting uncertainties for each bin are reflected by the accompanying vertical bars. {\bf Right column:} The upper panel shows the luminosity function for second-ranked LG galaxies and the lower panel shows the distribution of first-ranked galaxies, and the observational data is taken from Tago et al. (2008) groups (blue line) SDSS4\_Y08 (red line). Data here has been restricted to mimic a survey in which the magnitude gap between first- and fourth-ranked group galaxies is two.}

\label{fig:M1M2lim}
\end{figure*}

In order to make a fair comparison of the model predictions with the current observations, we take into account the selection effects inherent within the data.  Specifically, the observational results (i) have a limited dynamic range of $\sim$2~mag \citep{Lin1996},  driven by signal-to-noise constraints applied to the lowest luminosity galaxies in the survey, and (ii) discard groups that contain fewer than four galaxies within $\sim$2~mag of the first-ranked galaxy. \citet{Tago2008}  groups were chosen for the comparison because the absolute r-band magnitude data was readily available. We have imposed comparable selection effects upon the models, and the impact upon the luminosity functions of the first- and second-ranked CG galaxies is shown in the right hand panels of Fig. \ref{fig:M1M2lim}. The left panels show that the turnover in the Munich models is no longer apparent, once a dynamic range of two magnitudes is imposed upon the (theoretically, infinite) magnitude gap between first- and fourth-ranked group galaxies. With these cuts, the models and data now lie closer to one another. However, the models produce a significant shortage of pairs with low magnitude gaps for LGs and a higher population of groups with a magnitude gap of one. This effect is more extreme in the Munich models but is still present in D\_B06. 

What is perhaps more interesting is that the dynamic range need only be increased to three magnitudes for the models  to diverge significantly, with the M\_D06 model both ``broadening'' and shifting to lower luminosity, relative to the distributions based upon the M\_B07 and D\_B06 SAMs, (right-most column of Fig. \ref{fig:offset}). Table 4 shows the populations of groups in the two extreme cases of small and large magnitude gap for a dynamic range of 3~mag. The differences are very large between the Durham and Munich models. Certainly, observations with a higher dynamical range will provide a  good test for differentiating the success of the SAMs within group environments. The relative success of the different manner that the SAMs implement physical processes such as AGN and SN feedback, and how these become important within group environments where satellite accretion modelling is also crucial,  can then be better determined. One may also ask whether the assumption of separating Central and Satellite populations, whereby only the central galaxies experience mergers and no satellite galaxies grow while in the group environment, is appropriate when two galaxies of almost equal mass often exist within such environments.

The proportion of first-ranked (by luminosity) galaxies being centrals is sufficiently high to make the transition from the theoretical definitions of ``central'' and ``satellite'' galaxies into the observational regime of ``brightest'' and ``second brightest'' group galaxies - i.e., we can associate the brightest group galaxy with a central, and the second-brightest galaxy to a satellite. This then allows us to plot the luminosity function of first- (M1) and second-ranked (M2) group galaxies, as shown in the right hand panels of Fig.\ref{fig:M1M2lim}, and associate the distribution in M1 with model centrals, and the distribution in M2 with model (brightest) satellites. The first-ranked galaxy luminosity function of D\_B06 is broader and flatter than those of the two Munich SAM variants; as expected, the M\_B07 model galaxies are, on average more luminous. For the distribution of second-ranked galaxies, the M\_D06 galaxies are on average $\sim$1~mag less luminous than the Durham model galaxies, and the distribution is  broader. The right panels highlight that, although the global luminosity function of galaxies is well matched by observations, the distributions for the first and second ranked galaxies shown in Fig. \ref{fig:M1M2lim}  tend to be dimmer and wider than observations.

\begin{table}
\centering
  \caption{Percentage of groups with magnitude gap between first 
           and second ranked galaxies greater than 2
           magnitudes (top) and less than 0.5 magnitudes (bottom)} 
\label{table_gap}
\begin{tabular}{|c|c|c|c|c||}
  	&	 gap 	&	 D\_B06 	&	 M\_D06 	&	 M\_B07 	\\
\hline									
LG 	&	 $>2$   	&	10.8	&	20.6	&	15.2	\\
CG 	&	 $>2$   	&	14.9	&	28.4	&	23.8	\\
vCG  	&	 $>2$ 	&	18.2	&	27.5	&	29.8	\\
\hline									
LG 	&	 $<0.5$ 	&	35.6	&	22.7	&	28.7	\\
CG 	&	 $<0.5$ 	&	30.4	&	16.4	&	19.7	\\
vCG	&	 $<0.5$ 	&	27.5	&	17.8	&	18.7	\\
\hline									
\end{tabular}
\end{table}

\begin{figure*}
\centering
\includegraphics[trim = 0mm 0mm 0mm 0mm,clip,
  width=120mm]{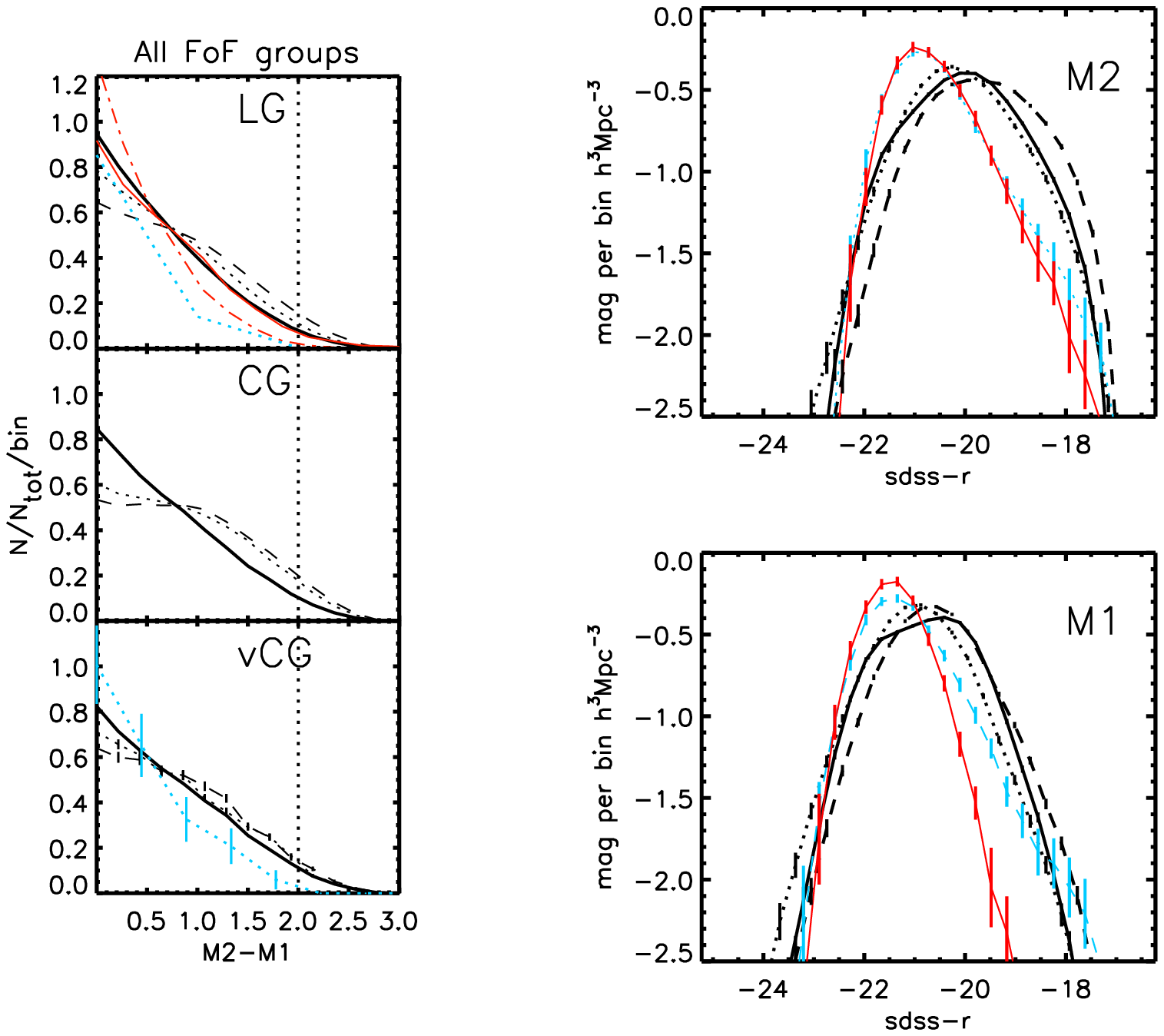} 
\caption{{\bf Left column:}  Distribution of the magnitude gap between 1st and 2nd ranked galaxies, normalised by total number of groups in each sample. The thick solid, dashed and dotted lines are for D\_B06, M\_D06, M\_B07 models. The coloured lines show observational data. The top left panel shows Tago et al. (2008) groups (red dashed line), SDSS4\_Y08 (red solid line) and Tucker et al. 2000 LGs (blue dotted line), the middle panel shows the results for CGs and and the bottom left panel shows Allam et al. (2000) compact groups (blue dotted line) along with vCGs. Poisson counting uncertainties for each bin are reflected by the accompanying vertical bars. {\bf Right column:} The upper panel shows the luminosity function for second-ranked LG galaxies and the lower panel shows the distribution of first-ranked galaxies, and the observational data is taken from Tago et al. (2008) groups (blue line) SDSS4\_Y08 (red line). In contrast with Fig. \ref{fig:M1M2lim}, data here has been restricted to mimic a survey in which the magnitude gap between first- and fourth-ranked group galaxies is three.}
 
\label{fig:M1M2lim2}
\end{figure*}

In Figure \ref{fig:M1M2lim2}, we demonstrate the impact of imposing a dynamic range of three magnitudes between first- and fourth-ranked group galaxies; having done so, we find that the observations of SDSS4\_Y08 match the model predictions of D\_B06 remarkably well for Loose Groups.  This suggests that the Durham model, in this regime, provides a better match to empirical data than that of the Munich models.

\section{Conclusions}

By constructing luminosity functions of galaxy groups (ranging from loose to very compact) using variants of several leading SAMs, as applied to the Millennium Simulation, we have explored an astrophysical regime in which the SAMs have not been inter-compared in great detail.  Several obvious differences between the M\_D06 \citep{DeLucia2006} and D\_B06 \citep{Bower2006} (i.e., loosely speaking, the Munich and Durham variants, respectively), became apparent, including an intermediate luminosity ``wiggle'' in the M\_D06  group luminosity functions not readily apparent when using the D\_B06 SAM.  We trace the origin of this wiggle to two competing effects resulting from the underlying physics within the M\_D06 SAM - a steeper faint-end slope to the satellite luminosity function and a narrower distribution to the central galaxies luminosity function, most likely due to the lack of mass stripping in satellite galaxies without enveloping subhaloes, type 2 groups, and the particular formulation of AGN in the Munich models. A systematic exploration of parameter space in the respective SAM may however, be required to further isolate the cause of the difference.

Observations suggest that such a wiggle in the group luminosity function might exist \citep{Weinmann2006}, similar to that seen when applying the M\_D06 SAM.  However, these same observations tend to show a steeper magnitude gap (between first- and second-ranked group members) distribution profile than that seen with any of SAMs, and we also see significant ``flattening'' in the M\_D06 gap distribution (i.e., a comparable likelihood for first- and second-ranked galaxies to be of equal luminosity, as to have a one magnitude luminosity difference), a feature that is not consistent with the data sets described by \citet{Miller2005} or \citet{Dariush2007}.

The models applied to the Millennium simulation produce noticeably different galaxy group properties. The group luminosity functions diverge with increasing galaxy density meaning that, for example, the cores of clusters in the various models have different properties, while the properties of the entire cluster will be more similar. As the same dark matter background was used in the three models there are similar numbers of groups and clusters in the models, but according to our definitions, the denser structures are several times more common in the D\_B06 model. The M\_D06 model luminosity function shows a peak for the brightest galaxies that does not appear in the Durham models and is less evident in M\_B07. The magnitude gap distributions of the models also differ with the Munich and Durham models demonstrating a different distribution at the small gap part of the distribution. All models show a shallower, wider magnitude gap distribution than the observations. This suggests that improvement in how the central / bright satellite luminosities are calculated is required. The designation of a single central galaxy which is modelled different manner to the other group members is a simplification which may need to be improved upon. 

The existence of more, denser, CG and vCG groups in the Durham models compared to the Munich sample  suggests that the different merging time-scales and implementations of satellite accretion can have noticeable effects on the predictions of the models. Similarly the fact that the Durham models show a shorter mean galaxy-galaxy separation, indicates that these groups are denser. This suggests that merging time-scales are longer in the Durham groups. This is backed up by the luminosity function of groups because the evident 'wiggle' in the M\_D06 groups appears to be due to a smaller population of satellites and brighter centrals, which is a direct result of the rate at which satellite galaxies are accreted onto the central galaxy. However, while observations show a similar `wiggle' in group luminosity functions, suggesting the shorter merging time is more physical, \citet{McConnachie2008} find fewer compact groups in their field than in the SAMs, suggesting the merging time-scale should be even shorter. Contrastingly the limited magnitude gap distribution indicates that the gap between central and satellite galaxies should be smaller, which may be due to additional physics that is not yet implemented in the models. Our analysis of the timescales of merging shows that this is not the case, as satellites in both models last a similar amount of time. We emphasises, however, that there are more galaxies near the centre of a given group/cluster in the Durham models, despite this. We suggest that this may be due to the additional time the M\_D06 model identifies subhaloes, and reassigns the galaxy position of the central galaxy as the most bound particle in the (sub)halo for longer. This may have the effect of keeping the galaxy out of the central region for longer. Although we do not find a noticeable difference in the merging times of galaxies in the two models there is a substantial population of groups which do not merge. We can see this because more haloes merge in the Durham model but more galaxies merge in the Munich models. This serves to build up the number of satellites in the cluster, which, fall into the cluster core, thus accounting for the observed difference in vCG population and galaxy density distribution. This can explain the difference in the magnitude gap distribution because more galaxies merge with the central in the Munich model, reducing the number of satellites and making the central galaxy brighter. 

\section*{Acknowledgements}
ONS acknowledges the support of the STFC through its PhD Studentship Programme. BKG and CBB acknowledge the support of the UK's Science \& Technology Facilities Council (STFC Grant ST/F002432/1) and the Commonwealth Cosmology Initiative; visitor support (PSB, DK, AK, LVS) from the STFC (ST/G003025/1) is similarly acknowledged. PSB acknowledges the support of a Marie Curie Intra-European Fellowship within the 6th European Community Framework Programme. AK and PSB are supported by the Ministerio de Ciencia e Innovacion (MICINN) in Spain through the Ramon y Cajal programme. The Millennium Simulation databases used in this paper and the web application providing on-line access to them were constructed as part of the activities of the German Astrophysical Virtual Observatory. Access to the University of Central Lancashire's High Performance Computing Facility is gratefully acknowledged. We acknowledge the computational support provided by the UK's National Cosmology Supercomputer, COSMOS. We thank the DEISA consortium, co-funded through EU FP6 project RI-031513 and the FP7 project RI-222919, for support within the DEISA Extreme Computing Initiative.

\bibliography{bib.bib}
\bibliographystyle{mn2e}

\end{document}